\RequirePackage{fix-cm}
\documentclass[smallextended]{svjour3}       
\smartqed  
\usepackage{amssymb}
\usepackage{amsmath}
\usepackage{color}
\usepackage{graphicx}
\usepackage[justification=centering]{caption}
\usepackage{multirow}
\begin{document}

\title{Improved Deterministic $N$-To-One Joint Remote Preparation of an Arbitrary Qubit via EPR Pairs
} \subtitle{}

\titlerunning{Improved Deterministic $N$-To-One Joint Remote Preparation of...}        

\author{Wen-Jie Liu       \and
        Zheng-Fei Chen    \and
        Chao Liu       \and
        Yu Zheng     
}


\institute{
           W.-J. Liu  \and
           Y. Zheng
           \at Jiangsu Engineering Center of Network Monitoring, Nanjing University of Information Science \& Technology, Nanjing 210044, P.R.China
           \\
           \email{wenjiel@163.com}
           \and
           W.-J. Liu   \and
           Z.-F Chen   \and
           C. Liu      \and
           Y. Zheng
           \at School of Computer and Software, Nanjing University of Information Science \& Technology, Nanjing 210044, P.R.China
}

\date{Received: date / Accepted: date}

\maketitle

\begin{abstract}
Recently, Bich et al. (Int. J. Theor. Phys. 51: 2272, 2012) proposed two deterministic joint remote state preparation (JRSP) protocols of an arbitrary single-qubit state: one is for two preparers to remotely prepare for a receiver by using two Einstein-Podolsky-Rosen (ERP) pairs; the other is its generalized form in the case of arbitrary $N$ ($N>2$) preparers via $N$ ERP pairs. While examining these two protocols, we find that the success probability for the receiver achieving the desired state is not deterministic, i.e., $P^{N>2}_{suc}<1$, for $N>2$ preparers in the second protocol. Through constructing two sets of adaptive projective measurement bases for both the real space and the complex space, an improved deterministic $N$-to-one JRSP protocol for an arbitrary single-qubit state is presented. Analysis shows our protocol can truly achieve the unit success probability, i.e., $P^{N\geq2}_{suc}=1$. What is more, the receiver can be randomly assigned even after the distribution of the qubits of EPR pairs, so it is more flexible and applicable in the network situation.

\end{abstract}

\section{Introduction}
\label{intro1}
With the rapid development of quantum mechanics in recent decades, many protocols of quantum information have been flourished by utilizing quantum mechanics principles, including quantum key distribution (QKD) [1-3], quantum secret sharing (QSS) [4-6], quantum direct communication (QDC) [7-9], quantum teleportation (QT) [10-12], quantum private comparison (QPC) [13-15], and so on. In the recent ten years, a new direction, remote state preparation (RSP) [16-18], has become a hot topic in the quantum information filed. Similar to QT, in RSP the preparer can exploit the nonlocal correlation of the quantum entangled state shared in advance to prepare the original state in the remote place. But the main difference is that the preparer must know all the information of the state in RSP, while in QT the preparer knows nothing about the state.

The early RSP protocols always focus on one preparer and one receiver, and the preparer knows all the information of the prepared state. However, for some highly sensitive and important information, it might be unreliable to let one preparer hold everything. To overcome this defection, joint remote state preparation (JRSP) was put forward. The pioneering JRSP protocol was proposed by Xia et al. [19] in 2007, in which the authors realize the multiparty remote preparation of the quantum state
$(\alpha|0\rangle^{\bigotimes M}+\beta|1\rangle^{\bigotimes M}, M=1,2,...,\infty)$
with unit fidelity but less than unit probability between $N-1$ senders and one receiver using one $N$-particle non-maximally entangled Greenberger-Horne-Zeilinger (GHZ) state as the quantum channel. Since then, JRSP has attracted a lot of attention in recent years [20-25]. Unfortunately, these protocols are probabilistic, i.e., they cannot be realized with unit success probability.

To ensure the unit success probability of JRSP, a new direction of JRSP, namely deterministic JRSP, has been put forward. In 2011, Xiao et al. [26] introduced a three-step strategy of JRSP to remotely prepare an arbitrary two- and three-qubit state by using GHZ states as the shared quantum resource. In this protocol, two preparers measure their qubits orderly rather than independently, and the success probability of preparation can be increased to 1. Inspired by the three-step strategy, many other deterministic JRSP protocols have been proposed [27-30]. For example, In 2012, An et al. [27] put forward a scheme to deterministically prepare the most general single- and two-qubit state by using four Einstein-Podolsky-Rosen (EPR) pairs as the shared quantum resource. And then, Chen et al. [28] proposed a scheme to joint remotely prepare an arbitrary three-qubit state deterministically by using six EPR pairs. In 2013, Wang et al. [29] proposed a scheme to jointly and remotely prepare an arbitrary two-qubit state, and generalize it to the arbitrary three-qubit case. Compared with Refs. [27] and [28], Wang et al.'s protocol just requires two EPR pairs and one GHZ state in the arbitrary two-qubit case and four EPR pairs and one GHZ states in the arbitrary three-qubit case. Being more robust and persistent than GHZ states, Cluster states are also utilized in JRSP. In 2013, Wang et al. [30] proposed a new version of deterministic JRSP protocol for an arbitrary two-qubit state by using the six-qubit cluster state. However, To the authors¡¯ knowledge, most of deterministic JRSP protocols only focus on the two-preparer case, and these protocols are impossible or difficult to be directly generalized to the arbitrary preparers case.

Recently, Bich et al. [31] proposed two deterministic JRSP protocols using ERP pairs: one is for two preparers to remotely prepare arbitrary single-qubit state; and the other is for $N$ ($N>2$) preparers. The authors claim that the total success probability is 1 ($P_{suc}=1$) both for two preparers and $N>2$ preparers. Unfortunately, we find it is not true in case of $N>2$ preparers: the success probability $P^{N>2}_{suc}<1$. In order to solve the problem, we tactfully constructed two sets of projective measurement bases, i.e., the real-coefficient measurement basis and the complex-coefficient measurement basis, and further proposed an $N$-to-one JRSP protocol for an arbitrary single-qubit state with unit success probability (i.e, $P^{N\geq2}_{suc}=1$).

The paper is organized as follows. In the next section, Bich et al.'s second protocols, especially the second protocol, are briefly reviewed and its correctness analysis is given in detail. In Sect. 3, two sets of projective measurement bases are constructed, and based on them, an improved deterministic $N$-to-one JRSP protocol of an arbitrary single-qubit state is proposed. Finally, a concise summary is rendered in Sect. 4.

\section{Review of Bich et al.'s protocols and the correctness analysis}
\label{sec:2}
\subsection{Review of Bich et al.'s protocols}
\label{sec:2.1}
In Bich et al.'s first protocol, they proposed a deterministic JRSP protocol wtih $N=2$ preparers and a receiver. In this protocol, two preparers, Alice and Bob, can help the remote receiver, Charlie, prepare a single-qubit state

\begin{equation}
|\phi\rangle=a|0\rangle+b^{i\varphi}|1\rangle,
\end{equation}
where coefficients $a,b$ are real with the normalization condition $|a|^{2}+|b|^{2}=1$, and $\varphi\in[0,2\pi]$. To our knowledge, the success probability for the receiver achieving the desired state is truly 1, so we just skip the detail of Bich et al.'s first protocol. In the appendix, Bich et al. propose the second protocol with $N>2$ preparers, in which the authors tried to directly generalize the first protocol to the situation of $N>2$ preparers, but unfortunately, there exists a fatal problem. Before reviewing the procedures of Bich et al.'s second protocol, we need have some knowledge about the pre-shared quantum channel among $N$ preparers (Alice, Bob 1, Bob 2, ..., Bob $N$) and a receiver (Charlie) as shown in Fig. 1.
\begin{figure}
  \centering
  \includegraphics[width=8.5cm]{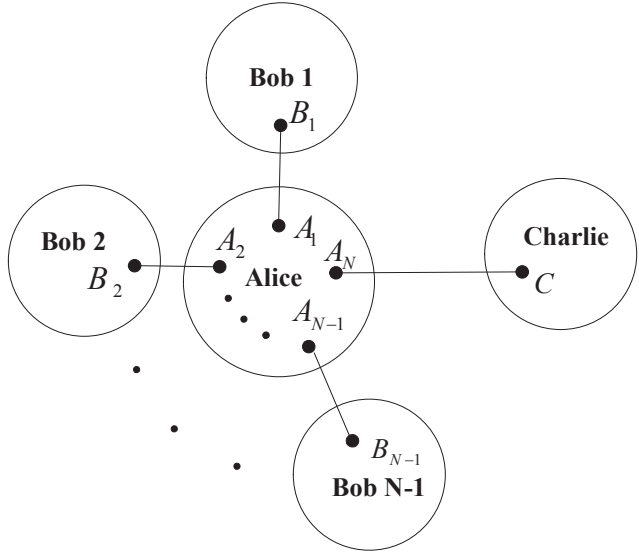}\\
  \caption{$^{[31]}$ The qubits¡¯ distribution for JRSP of the most general single-qubit state via $N$ EPR pairs for the situation of $N$ preparers (Alice, Bob 1, Bob 2, . . . and Bob N-1) and a receiver (Charlie). Qubits are represented by dots and entangled qubits are connected by a solid line.}\label{Fig. 1}
\end{figure}

For the convenience of description, the authors took $N=3$ (Alice, Bob 1, and Bob 2), and the participants pre-share three EPR pairs as the quantum channel
\begin{equation}
|Q\rangle_{A_{1}B_{1}A_{2}B_{2}A_{3}C}=|EPR\rangle_{A_{1}B_{1}}|EPR\rangle_{A_{2}B_{2}}|EPR\rangle_{A_{3}C},
\end{equation}
where $|EPR\rangle=\frac{1}{\sqrt{2}}(|00\rangle+|11\rangle)$, qubits $A_{1},A_{2},A_{3}$ are hold by Alice, and qubits $B_{1},B_{2},C$ by Bob 1, Bob 2 and Charlie, respectively. Alice is allowed to know $\{a,b\}$ , Bob 1 and Bob 2 share the knowledge of $\varphi$ in the following way: Bob 1 knows $\varphi_{1}$ and Bob 2 knows $\varphi_{2}$, and $\varphi=\varphi_{1}+\varphi_{2}$. The detailed three-step protocol for $N=3$ preparers can be described as below.

\textbf{Step 1}. Alice measures her three qubits in the basis $\{|u_{klm}\rangle_{A_{1}A_{2}A_{3}};k,l,m\in\{0,1\}\}$,
\begin{equation}
\left(
  \begin{array}{c}
    |u_{000}\rangle_{A_{1}A_{2}A_{3}} \\
    |u_{001}\rangle_{A_{1}A_{2}A_{3}} \\
    |u_{010}\rangle_{A_{1}A_{2}A_{3}} \\
    |u_{011}\rangle_{A_{1}A_{2}A_{3}} \\
    |u_{100}\rangle_{A_{1}A_{2}A_{3}} \\
    |u_{101}\rangle_{A_{1}A_{2}A_{3}} \\
    |u_{110}\rangle_{A_{1}A_{2}A_{3}} \\
    |u_{111}\rangle_{A_{1}A_{2}A_{3}} \\
  \end{array}
\right)
=\left(
   \begin{array}{cccccccc}
     a & ~~~0 & ~~~0 & ~~0 & ~~0 & ~~0 & ~0 & ~b \\
     b & ~~~0 & ~~~0 & ~~0 & ~~0 & ~~0 & ~0 & -a \\
     0 & ~~~a & ~~~0 & ~~0 & ~~0 & ~~0 & ~b & ~0 \\
     0 & ~~~b & ~~~0 & ~~0 & ~~0 & ~~0 & -a & ~0 \\
     0 & ~~~0 & ~~~a & ~~0 & ~~0 & ~~b & ~0 & ~0 \\
     0 & ~~~0 & ~~~b & ~~0 & ~~0 & ~-a & ~0 & ~0 \\
     0 & ~~~0 & ~~~0 & ~~a & ~~b & ~~0 & ~0 & ~0 \\
     0 & ~~~0 & ~~~0 & ~-b & ~~a & ~~0 & ~0 & ~0 \\
   \end{array}
 \right)
\left(
  \begin{array}{c}
    |000\rangle_{A_{1}A_{2}A_{3}} \\
    |001\rangle_{A_{1}A_{2}A_{3}} \\
    |010\rangle_{A_{1}A_{2}A_{3}} \\
    |011\rangle_{A_{1}A_{2}A_{3}} \\
    |100\rangle_{A_{1}A_{2}A_{3}} \\
    |101\rangle_{A_{1}A_{2}A_{3}} \\
    |110\rangle_{A_{1}A_{2}A_{3}} \\
    |111\rangle_{A_{1}A_{2}A_{3}} \\
  \end{array}
\right),
\end{equation}
and publicly broadcasts the measurement outcome to Bob 1, Bob 2 and Charlie. Expressing the quantum channel $|Q\rangle_{A_{1}B_{1}A_{2}B_{2}A_{3}C}$ through $|u_{klm}\rangle_{A_{1}A_{2}A_{3}}$,
\begin{equation}
{\left| Q \right\rangle _{{A_1}{B_1}{A_2}{B_2}{A_3}C}} = \frac{1}{{2\sqrt 2 }}\sum\limits_{m = 0}^1 {\sum\limits_{l = 0}^1 {\sum\limits_{k = 0}^1 {{{\left| {{u_{klm}}} \right\rangle }_{{A_1}{A_2}{A_3}}}{{\left| {{L_{klm}}} \right\rangle }_{{B_1}{B_2}C}}} } },
\end{equation}
we derive $|L_{klm}\rangle_{B_{1}B_{2}C}$ in the form
\begin{equation}
|L_{000}\rangle_{B_{1}B_{2}C}=a|000\rangle_{B_{1}B_{2}C}+b|111\rangle_{B_{1}B_{2}C},~~
\end{equation}
\begin{equation}
|L_{001}\rangle_{B_{1}B_{2}C}=-a|111\rangle_{B_{1}B_{2}C}+b|000\rangle_{B_{1}B_{2}C},
\end{equation}
\begin{equation}
|L_{010}\rangle_{B_{1}B_{2}C}=a|001\rangle_{B_{1}B_{2}C}+b|110\rangle_{B_{1}B_{2}C},~~
\end{equation}
\begin{equation}
|L_{011}\rangle_{B_{1}B_{2}C}=-a|110\rangle_{B_{1}B_{2}C}+b|001\rangle_{B_{1}B_{2}C},
\end{equation}
\begin{equation}
|L_{100}\rangle_{B_{1}B_{2}C}=a|010\rangle_{B_{1}B_{2}C}+b|101\rangle_{B_{1}B_{2}C},~~
\end{equation}
\begin{equation}
|L_{101}\rangle_{B_{1}B_{2}C}=-a|101\rangle_{B_{1}B_{2}C}+b|010\rangle_{B_{1}B_{2}C},
\end{equation}
\begin{equation}
|L_{110}\rangle_{B_{1}B_{2}C}=a|011\rangle_{B_{1}B_{2}C}+b|100\rangle_{B_{1}B_{2}C},~~
\end{equation}
\begin{equation}
|L_{111}\rangle_{B_{1}B_{2}C}=a|100\rangle_{B_{1}B_{2}C}-b|011\rangle_{B_{1}B_{2}C}.~~
\end{equation}

\textbf{Step 2}. Bob 1 and Bob 2 independently measure their qubits. If $klm=000$ or $010$, each Bob $j$ ($j=1,2$) uses a measurement basis determined by
\begin{equation}
\left( \begin{array}{l}
{\left| {{v_0}} \right\rangle _{{B_j}}}\\
{\left| {{v_1}} \right\rangle _{{B_j}}}
\end{array} \right) = {V^{\left( 0 \right)}}\left( {{\varphi _j}} \right)\left( \begin{array}{l}
{\left| 0 \right\rangle _{{B_j}}}\\
{\left| 1 \right\rangle _{{B_j}}}
\end{array} \right);
\end{equation}
If $klm=001$ or $011$, the basis for each Bob $j$ is
\begin{equation}
\left( \begin{array}{l}
{\left| {{v_0}} \right\rangle _{{B_j}}}\\
{\left| {{v_1}} \right\rangle _{{B_j}}}
\end{array} \right) = {V^{\left( 1 \right)}}\left( {{\varphi _j}} \right)\left( \begin{array}{l}
{\left| 0 \right\rangle _{{B_j}}}\\
{\left| 1 \right\rangle _{{B_j}}}
\end{array} \right),
\end{equation}
where
\begin{equation}
{V^{(r)}}\left( \varphi  \right) = \frac{1}{{\sqrt 2 }}\left( {\begin{array}{*{20}{c}}
1&~~~{{e^{ - {{\left( { - 1} \right)}^r}i\varphi }}}\\
{{e^{{{\left( { - 1} \right)}^r}i\varphi }}}&~~~{ - 1}
\end{array}} \right),r \in \left\{ {0,1} \right\}.
\end{equation}
In case $klm=100$ or $110$, Bob 1 uses a measurement basis determined by Eq.(13), and the basis for Bob 2 is Eq. (14); while $klm=101$ or $111$, the bases for Bob 1 and Bob 2 are determined by Eq. (14) and (13), respectively. After measurement, Bob 1 and Bob 2 announce their measurement outcomes $n,s$ ($n,s\in\{0,1\}$) to Charlie.

\textbf{Step 3}. Charlie converts the qubit $C$ to be the desired state $|\phi\rangle_{C}$ by applying reconstruction operators $R_{klmns}$. The operators $R_{klmns}$ that Charlie needs in this step are shown in Table 1.

\label{tab:1}
\begin{table}[h]\normalsize
\caption{$^{[31]}$ The reconstruction operator $R_{klmns}$, conditioned on the measurement outcomes $klm$, $n$ and $s$ of Alice, Bob 1 and Bob 2, respectively. $I$ is the identity operator, $X=|0\rangle\langle1|+|1\rangle\langle0|$ the bit-flip operator and $Z=|0\rangle\langle0|-|1\rangle\langle1|$ the phase-flip one}
\begin{tabular}{lllllllllllll}
\hline\noalign{\smallskip}
$\sharp$ & $klmns$ & $R_{klmns}$\\
\noalign{\smallskip}\hline\noalign{\smallskip}
1-8 & $00000,00011,10000,10011,01101,01110,11101 or 11100$ & $I$\\
9-16 & $01000,01011,11000,11011,00101,00110,10101 or 10110$ & $X$\\
17-24 & $00100,00111,10100,10111,01001,01010,11001123 or 11010$ & $ZX$\\
25-32 & $01100,01111,11100,11111,00001,00010,10001 or 10010$ & $Z$\\
\noalign{\smallskip}\hline
\end{tabular}
\end{table}

\subsection{Correctness analysis}
\label{sec:2.2}
In Ref. [31], the authors claimed the receiver can get the desirable state with the unit success probability ($P_{suc}=1$) both for two preparers and $N>2$ preparers. In fact, it is not true in the case of $N>2$ preparers. In the following, we will analyze its correctness in detail.

In Step 1, Alice measures her three qubits in the basis $\{|u_{klm}\rangle_{A_{1}A_{2}A_{3}};k,l,m\in\{0,1\}\}$ and publishes her outcome to the other participants (Bob 1, Bob 2 and Charlie). For sake of clearness, we assume the measurement outcome is $klm=010$, and according to the selection strategy of measurement basis in Step 2, Bob 1 and Bob 2 will use the measurement basis of Equation (13), i.e., the basis for Bob 1 is
$\left(
\begin{array}{c}
|v_{0}\rangle_{B_{1}} \\
|v_{1}\rangle_{B_{1}}\\
\end{array}
\right)
=V^{0}(\varphi_{1})\left(
\begin{array}{c}
|0\rangle_{B_{1}} \\
|1\rangle_{B_{1}}\\
\end{array}
\right)=\frac{1}{{\sqrt 2 }}\left(
\begin{array}{cc}
1 & e^{-i\varphi_{1}} \\
e^{i\varphi_{1}} & -1 \\
\end{array}
\right)\left(
\begin{array}{c}
|0\rangle_{B_{1}} \\
|1\rangle_{B_{1}}\\
\end{array}
\right)$,
and the basis for Bob  2 is $\left(
\begin{array}{c}
|v_{0}\rangle_{B_{2}} \\
|v_{1}\rangle_{B_{2}}\\
\end{array}
\right)
=V^{0}(\varphi_{2})\left(
\begin{array}{c}
|0\rangle_{B_{2}} \\
|1\rangle_{B_{2}}\\
\end{array}
\right)=\frac{1}{{\sqrt 2 }}\left(
\begin{array}{cc}
1 & e^{-i\varphi_{2}} \\
e^{i\varphi_{2}} & -1 \\
\end{array}
\right)\left(
\begin{array}{c}
|0\rangle_{B_{2}} \\
|1\rangle_{B_{2}}\\
\end{array}
\right)$. Then, $|L_{010}\rangle_{B_{1}B_{2}C}$ can be rewritten as

\begin{equation}
\begin{split}
|L_{010}\rangle_{B_{1}B_{2}C}=&\frac{1}{2}\sum^{1}_{n=0}\sum^{1}_{s=0}|v_{n}\rangle_{B_{1}}|v_{s}\rangle_{B_{2}}|\phi\rangle_{C}\\
=&\frac{1}{2}|v_{0}\rangle_{B_{1}}|v_{0}\rangle_{B_{2}}(a|1\rangle+be^{i(\varphi_{1}+\varphi_{2})}|0\rangle)_{C}\\
+&\frac{1}{2}|v_{0}\rangle_{B_{1}}|v_{1}\rangle_{B_{2}}(ae^{-i\varphi_{2}}|1\rangle-be^{i\varphi_{1}}|0\rangle)_{C}\\
+&\frac{1}{2}|v_{1}\rangle_{B_{1}}|v_{0}\rangle_{B_{2}}(ae^{-i\varphi_{1}}|1\rangle-be^{i\varphi_{2}}|0\rangle)_{C}\\
+&\frac{1}{2}|v_{1}\rangle_{B_{1}}|v_{1}\rangle_{B_{2}}(ae^{-i(\varphi_{1}+\varphi_{2})}|1\rangle+b|0\rangle)_{C}.\\
\end{split}
\end{equation}
Obviously, only when $ns=00$, qubit $C$ will collapse into $|\phi\rangle_{C}=a|1\rangle+be^{i(\varphi_{1}+\varphi_{2})}|0\rangle$, and the receiver can achieve the desirable state just with the identity operator. But if $ns=01$ ($|\phi\rangle_{C}=ae^{-i\varphi_{2}}|1\rangle-be^{i\varphi_{1}}|0\rangle$), $10$ ($|\phi\rangle_{C}=ae^{-i\varphi_{1}}|1\rangle-be^{i\varphi_{2}}|0\rangle$) or $11$ ($|\phi\rangle_{C}=ae^{-i(\varphi_{1}+\varphi_{2})}|1\rangle+b|0\rangle$), there is no any appropriate unitary operator that can be adopted to transform the state $|\phi\rangle_{C}$ to the desirable state. That is to say, the probability for Charlie finding the desirable state is 1/4 under this situation.

In the general case, there are eight possible outcomes $\{000,001,010,011,100,\\101,110,111\}$ for Alice, and the probability of each outcome is equal. That means, the probability for Alice obtaining each state $|u_{klm}\rangle_{A_{1}A_{2}A_{3}}$ ($k,l,m=0,1$) is 1/8. Similar to the case of $klm=010$, Charlie can also obtain the desirable state with the probability of 1/4 in the other seven cases. So, the total success probability is
\begin{equation}
{P^{N=3}_{suc}} = \left( {\frac{1}{8} \times \frac{1}{4}} \right) \times 8 = \frac{1}{4} < 1.
\end{equation}
Clearly, this is contrary to what the authors claimed: ``the total success probability is obviously 1" [31].

\section{Improved deterministic $N$-to-one JRSP of an arbitrary qubit}
\label{sec:3}
\subsection{The construction of measurement basis}
\label{sec:3.1}
The construction of measurement basis is at the heart of RSP or JRSP, and it is the key point that will influence the final success probability of obtaining the desirable state. In order to improve the success probability of the $N$-to-one JRSP to 1, the first task is to tactfully construct two suitable sets of measurement bases, the real-coefficient measurement basis and the complex-coefficient one.

In our JRSP protocol, we will use the projective measurements to measure the qubits. Before constructing these ingenious projective measurement bases, we need have some knowledge of projective measurement.

\textbf{Theorem 1} \emph{Projective measurement}$^{[32]}$: A projective measurement is described by an observable, $M$, a Hermitian operator on the state space of the system being observed. The observable has a spectral decomposition,
\begin{equation}
M{\rm{ = }}\sum\limits_m {m{P_m}},
\end{equation}
where $P_{m}$ is the projector onto the eigenspace of $M$ with eigenvalue $m$. The possible outcomes of the measurement correspond to the eigenvalues, $m$, of the observable.

According to the definition of projective measurement (Theorem 1), projective measurement can be understood as a special case of Postulate 3 (i.e., the quantum measurement postulate, seen in Refs. [32] and [33]). Specifically, a complete set of orthogonal projectors ${P_m}$ satisfy the following relations: (1) $\sum\limits_m {P_m} = I$; and (2) ${P_m}{P_{m'}} = {\delta _{m,m'}}{P_m}$ (${\delta _{m,m'}}$ = 1, if ${m = m'}$; ${\delta _{m,m'}}$ = 0, if ${m \ne m'}$). In quantum information field, a widely used phrase, to `measure in a basis $\left| m \right\rangle$', where $\left| m \right\rangle$ form an orthonormal basis, simply means to perform the projective measurement with projectors as below
\begin{equation}
{P_m}{\rm{ = }}\left| m \right\rangle \left\langle m \right|.
\end{equation}

In our protocol, Alice holds $N$ qubits $A_{1},A_{2},...A_{N}$, and she knows the real part \{a,b\} of the state (given in Eq. (1)). So the projection operators $\{P_{m}\}$ can be described as ${P_m} = {\left| {{u_{{l_1},{l_2}, \ldots ,{l_N}}}} \right\rangle _{{A_1},{A_2}, \ldots ,{A_N}}}{\left\langle {{u_{{l_1},{l_2}, \ldots ,{l_N}}}} \right|_{{A_1},{A_2}, \ldots ,{A_N}}}$ ($l_{1},l_{2},...l_{N}\in\{0,1\}$), where the set of states $\{|u_{l_{1},l_{2},...l_{N}}\rangle_{A_{1},A_{2},...A_{N}}\}$ is a complete set of orthonormal basis and can be defined as
\begin{equation}
\left( {\begin{array}{*{20}{c}}
{\left| {{u_{00 \ldots 0}}} \right\rangle }\\
{\left| {{u_{00 \ldots 1}}} \right\rangle }\\
{}\\
{\begin{array}{*{20}{c}}
 \vdots \\
{}\\
{\left| {{u_{11 \ldots 0}}} \right\rangle }\\
{\left| {{u_{11 \ldots 1}}} \right\rangle }
\end{array}}
\end{array}} \right) = \left( {\begin{array}{*{20}{c}}
{\begin{array}{*{20}{c}}
a&~~~0\\
0&~~~a
\end{array}}& \ldots &{\begin{array}{*{20}{c}}
~0&~~~b\\
~b&~~~0
\end{array}}\\
 \vdots &{\begin{array}{*{20}{c}}
 \ddots &{}& {\mathinner{\mkern2mu\raise1pt\hbox{.}\mkern2mu
 \raise4pt\hbox{.}\mkern2mu\raise7pt\hbox{.}\mkern1mu}} \\
{}&{\begin{array}{*{20}{c}}
a&~~~b\\
{ - b}&~~~a
\end{array}}&{}\\
 {\mathinner{\mkern2mu\raise1pt\hbox{.}\mkern2mu
 \raise4pt\hbox{.}\mkern2mu\raise7pt\hbox{.}\mkern1mu}} &{}& \ddots
\end{array}}& \vdots \\
{\begin{array}{*{20}{c}}
0&~~~{ - b}\\
{ - b}&~~~0
\end{array}}& \cdots &{\begin{array}{*{20}{c}}
~a&~~~0\\
~0&~~~a
\end{array}}
\end{array}} \right)\left( {\begin{array}{*{20}{c}}
{\left| {00...0} \right\rangle }\\
{\left| {00...1} \right\rangle }\\
{}\\
{\begin{array}{*{20}{c}}
 \vdots \\
{}\\
{\left| {11...0} \right\rangle }\\
{\left| {11...1} \right\rangle }
\end{array}}
\end{array}} \right).
\end{equation}
The Eq. (20) can be generalized to a collection $\{|u_{l_{1}l_{2}...l_{N}}\rangle=a|l_{1}l_{2}...l_{N}\rangle_{A_{1},A_{2},...A_{N}}\\+
(-1)^{l_{1}}b|\overline{l_{1}}\overline{l_{2}}...\overline{l_{N}}\rangle_{A_{1},A_{2},...A_{N}}
{\kern 2pt}|{\kern 2pt}l_{k}\in\{0,1\},\overline{l_{k}}=1-l_{k},1\leq k \leq N\}$. Simultaneously, since each Bob $j$ ($1\leq j\leq N-1$) holds one qubit $B_{j}$ and knows a part of complex part $\varphi_{j}$, so, the corresponding complete set of orthonormal basis can be expressed as
\begin{equation}
\left( {\begin{array}{*{20}{c}}
{{{\left| {{v_0}} \right\rangle }_{{B_j}}}}\\
{{{\left| {{v_1}} \right\rangle }_{{B_j}}}}
\end{array}} \right) = {V^{'\left( {{l_j}} \right)}}\left( {{\varphi _j}} \right)\left( {\begin{array}{*{20}{c}}
{{{\left| 0 \right\rangle }_{{B_j}}}}\\
{{{\left| 1 \right\rangle }_{{B_j}}}}
\end{array}} \right),
\end{equation}
where $l_{j}$ represents the value of subscript in $|u_{l_{1}l_{2}...l_{N}}\rangle$, $l_{j}\in\{0,1\}$, and
\begin{equation}
{V^{'\left( 0 \right)}}\left( {{\varphi _j}} \right) = \frac{1}{{\sqrt 2 }}\left( {\begin{array}{*{20}{c}}
1&{{e^{ - i{\varphi _j}}}}\\
1&{ - {e^{ - i{\varphi _j}}}}
\end{array}} \right),
\end{equation}
\begin{equation}
{V^{'\left( 1 \right)}}\left( {{\varphi _j}} \right) = \frac{1}{{\sqrt 2 }}\left( {\begin{array}{*{20}{c}}
{{e^{ - i{\varphi _j}}}}&~~1\\
{ - {e^{ - i{\varphi _j}}}}&~~1
\end{array}} \right).
\end{equation}

\subsection{Our $N$-To-One JRSP protocol}
\label{sec:3.2}
By utilizing the measurement bases constructed in Sect. 3.1, we present an improved deterministic $N$-to-one JRSP of an arbitrary qubit as below.

\textbf{Prerequisite}. Suppose preparers (Alice, Bob 1, Bob 2, \ldots and Bob N-1) want to help the receiver Charlie prepare an arbitrary single-qubit state (given in Eq. (1)), and they share $N$ EPR pairs as quantum channel, which is described as
\begin{equation}
|Q\rangle_{A_{1}B_{1}A_{2}B_{2}...A_{N}C}=|EPR\rangle_{A_{1}B_{1}}|EPR\rangle_{A_{2}B_{2}}...|EPR\rangle_{A_{N-1}B_{N-1}}|EPR\rangle_{A_{N}C},
\end{equation}
Here, Alice is allowed to know $\{a,b\}$, Bob 1, Bob 2, . . . and Bob N-1 share the knowledge of $\varphi$, and each Bob $j$ ($1\leq j\leq N-1$) knows $\varphi_{j}$, where $\varphi  = \sum\limits_{j = 1}^{N - 1} {{\varphi _j}}$.

\textbf{Step 1}. Alice measures her $N$ qubits with the basis (given in Eq. (20)) and publicly broadcasts the measuring outcome $l_{1}l_{2}...l_{N}$ ($l_{1},l_{2},...l_{N}\in\{0,1\}$) to Bob 1, Bob 2, . . . Bob N-1 and Charlie.

\textbf{Step 2}. Each Bob $j$ independently measures her qubit $B_{j}$ in the basis conditioned on $\varphi_{j}$ and Alice's outcome $l_{j}$. Specifically, each Bob $j$ uses a basic measurement shown in Eq. (21). After measurement, each Bob $j$ announces her measurement outcome $m_{j}$ $(m_{j}\in\{0,1\})$ to Charlie.

\textbf{Step 3}. Charlie converts the qubit $C$ to be the desired state $|\phi\rangle_{C}$ by applying reconstruction operators $R_{\underbrace {l_{1}l_{2}...l_{N}}_N\underbrace {m_{1}m_{2}...m_{N-1}}_{N - 1}} = Z^{l_{N}\oplus\underbrace {m_{1} \oplus m_{2}\oplus ... \oplus m_{N-1}}_{N - 1}}\\X^{\underbrace {l_{1} \oplus l_{2}\oplus ... \oplus l_{N}}_{N}}$.

 The whole process of our $N$-to-one JRSP protocol can also be described in a quantum circuit form as shown in Fig. 2. For clarity, we consider $N=3$ (Alice, Bob 1, Bob 2) as example, and $|Q\rangle_{A_{1}B_{1}A_{2}B_{2}A_{3}C}=|EPR\rangle_{A_{1}B_{1}}|EPR\rangle_{A_{2}B_{2}}|EPR\rangle_{A_{3}C}$ is the quantum channel. At first, Alice measures her three qubits in the basis $\{|u_{l_{1}l_{2}l_{3}}\rangle_{A_{1}A_{2}A_{3}};l_{1},l_{2},l_{3}\in\{0,1\}\}$:
\begin{equation}
\left(
  \begin{array}{c}
    |u_{000}\rangle_{A_{1}A_{2}A_{3}} \\
    |u_{001}\rangle_{A_{1}A_{2}A_{3}} \\
    |u_{010}\rangle_{A_{1}A_{2}A_{3}} \\
    |u_{011}\rangle_{A_{1}A_{2}A_{3}} \\
    |u_{100}\rangle_{A_{1}A_{2}A_{3}} \\
    |u_{101}\rangle_{A_{1}A_{2}A_{3}} \\
    |u_{110}\rangle_{A_{1}A_{2}A_{3}} \\
    |u_{111}\rangle_{A_{1}A_{2}A_{3}} \\
  \end{array}
\right)
=\left(
   \begin{array}{cccccccc}
     a & ~0 & ~0 & ~0 & ~~0 & ~~0 & ~0 & ~b \\
     0 & ~a & ~0 & ~0 & ~~0 & ~~0 & ~b & ~0 \\
     0 & ~0 & ~a & ~0 & ~~0 & ~~b & ~0 & ~0 \\
     0 & ~0 & ~0 & ~a & ~~b & ~~0 & ~0 & ~0 \\
     0 & ~0 & ~0 & -b & ~~a & ~~b & ~0 & ~0 \\
     0 & ~0 & -b & ~0 & ~~0 & ~~a & ~0 & ~0 \\
     0 & -b & ~0 & ~0 & ~~0 & ~~0 & ~a & ~0 \\
     -b & ~0 & ~0 & ~0 & ~~0 & ~~0 & ~0 & ~a \\
   \end{array}
 \right)
\left(
  \begin{array}{c}
    |000\rangle_{A_{1}A_{2}A_{3}} \\
    |001\rangle_{A_{1}A_{2}A_{3}} \\
    |010\rangle_{A_{1}A_{2}A_{3}} \\
    |011\rangle_{A_{1}A_{2}A_{3}} \\
    |100\rangle_{A_{1}A_{2}A_{3}} \\
    |101\rangle_{A_{1}A_{2}A_{3}} \\
    |110\rangle_{A_{1}A_{2}A_{3}} \\
    |111\rangle_{A_{1}A_{2}A_{3}} \\
  \end{array}
\right),
\end{equation}

\begin{figure}
  \centering
  \includegraphics[width=10.5cm]{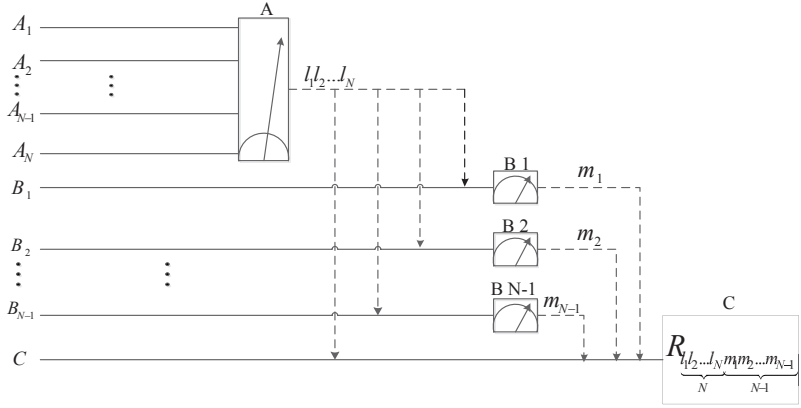}\\
  \caption{The quantum circuit for deterministic deterministic $N$-to-one JRSP of an arbitrary qubit via $N$ EPR pairs. Particles $A_{1}A_{2}...A_{N}$ are hold by Alice, $B_{1},B_{2},...,B_{N-1}$ hold by Bob 1, Bob 2, . . . and Bob N-1, respectively, and $C$ hold by Charlie. The classical communications are represented as the dashed arrows, while the quantum channels are represented as the solid lines. what is more, boxes A, B1, B2,¡­ and B N-1 depict Alice$'$s, Bob 1$'$s, Bob 2$'$s, ..., and Bob N-1$'$s measurement operation, respectively, and the last box $C$ denotes the unitary operation Charlie carries out.} \label{Fig. 2}
\end{figure}
and publicly broadcasts outcome $l_{1}l_{2}l_{3}$. Before measurement, the quantum channel can be rewritten as
\begin{equation}
{\left| Q \right\rangle _{{A_1}{B_1}{A_2}{B_2}{A_3}C}} = \frac{1}{{2\sqrt 2 }}\sum\limits_{{l_1} = 0}^1 {\sum\limits_{{l_2} = 0}^1 {\sum\limits_{{l_3} = 0}^1 {{{\left| {{u_{{l_1}{l_2}{l_3}}}} \right\rangle }_{_{{A_1}{A_2}{A_3}}}}{{\left| {{L_{{l_1}{l_2}{l_3}}}} \right\rangle }_{_{{B_1}{B_2}C}}}} } }.
\end{equation}
And we derive $|L_{l_{1}l_{2}l_{3}}\rangle_{B_{1}B_{2}C}$ in the form
\begin{equation}
|L_{l_{1}l_{2}l_{3}}\rangle_{B_{1}B_{2}C}=a|l_{1}l_{2}l_{3}\rangle_{B_{1}B_{2}C}+(-1)^{l_{1}}b|\overline{l_{1}}\overline{l_{2}}\overline{l_{3}}\rangle_{B_{1}B_{2}C}.
\end{equation}

After Alice's measurement, Bob 1 and Bob 2 use the basis to measure their own qubit. The state in Eq. (27) can be expressed as
\begin{equation}
|L_{l_{1}l_{2}l_{3}}\rangle_{B_{1}B_{2}C}= \frac{1}{2}\sum\limits_{{m_1} = 1}^2 {\sum\limits_{{m_2} = 1}^2 {} } {\left| {{v_{{m_1}}}} \right\rangle _{{B_1}}}{\left| {{v_{{m_2}}}} \right\rangle _{{B_2}}}|\phi\rangle _C.
\end{equation}
We assume $l_{1}l_{2}l_{3}=010$ without generality, then Bob 1 and Bob 2 choose the basis $\{|v_{m_{1}}\rangle_{B_{1}}~|m_{1}\in\{0,1\}\}$£º
\begin{equation}
\left( {\begin{array}{*{20}{c}}
{{{\left| {{v_0}} \right\rangle }_{{B_1}}}}\\
{{{\left| {{v_1}} \right\rangle }_{{B_1}}}}
\end{array}} \right) = \frac{1}{{\sqrt 2 }}\left( {\begin{array}{*{20}{c}}
1&~~{{e^{ - i{\varphi _1}}}}\\
1&~~{ - {e^{ - i{\varphi _1}}}}
\end{array}} \right)\left( {\begin{array}{*{20}{c}}
{{{\left| 0 \right\rangle }_{{B_1}}}}\\
{{{\left| 1 \right\rangle }_{{B_1}}}}
\end{array}} \right),
\end{equation}
and $\{|v_{m_{2}}\rangle_{B_{2}}~|m_{2}\in\{0,1\}\}$:
\begin{equation}
\left( {\begin{array}{*{20}{c}}
{{{\left| {{v_0}} \right\rangle }_{{B_2}}}}\\
{{{\left| {{v_1}} \right\rangle }_{{B_2}}}}
\end{array}} \right) = \frac{1}{{\sqrt 2 }}\left( {\begin{array}{*{20}{c}}
{{e^{ - i{\varphi _2}}}}&~~1\\
{ - {e^{ - i{\varphi _2}}}}&~~1
\end{array}} \right)\left( {\begin{array}{*{20}{c}}
{{{\left| 0 \right\rangle }_{{B_2}}}}\\
{{{\left| 1 \right\rangle }_{{B_2}}}}
\end{array}} \right),
\end{equation}
respectively. We can rewritten $|L_{010}\rangle_{B_{1}B_{2}C}$ as
\begin{equation}
\begin{split}
|L_{010}\rangle_{B_{1}B_{2}C}=&\frac{1}{2}a(|v_{0}\rangle_{B_{1}}+|v_{1}\rangle_{B_{1}})(|v_{0}\rangle_{B_{2}}+|v_{1}\rangle_{B_{2}})|0\rangle_{C}\\
+&\frac{1}{2}be^{i(\varphi_{1}+\varphi_{2})}(|v_{0}\rangle_{B_{1}}-|v_{1}\rangle_{B_{1}})(|v_{0}\rangle_{B_{2}}-|v_{1}\rangle_{B_{2}})|1\rangle_{C}\\
=&\frac{1}{2}|v_{0}\rangle_{B_{1}}|v_{0}\rangle_{B_{2}}(a|0\rangle+be^{i(\varphi_{1}+\varphi_{2})}|1\rangle)_{C}\\
+&\frac{1}{2}|v_{0}\rangle_{B_{1}}|v_{1}\rangle_{B_{2}}(a|0\rangle-be^{i(\varphi_{1}+\varphi_{2})}|1\rangle)_{C}\\
+&\frac{1}{2}|v_{1}\rangle_{B_{1}}|v_{0}\rangle_{B_{2}}(a|0\rangle-be^{i(\varphi_{1}+\varphi_{2})}|1\rangle)_{C}\\
+&\frac{1}{2}|v_{1}\rangle_{B_{1}}|v_{1}\rangle_{B_{2}}(a|0\rangle+be^{i(\varphi_{1}+\varphi_{2})}|1\rangle)_{C}.\\
\end{split}
\end{equation}
\begin{table}[h]\normalsize\centering
\caption{The collapsed state $|\phi\rangle_{C}$ of Charlie's qubits $C$ and the corresponding recovery operators $R_{l_{1}l_{2}l_{3}m_{1}m_{2}}$ related to the measurement outcomes ($l_{1}l_{2}l_{3},m_{1},m_{2}$) from Alice, Bob 1 and Bob 2}
\label{tab:2}
\begin{tabular}{cccccccc}
\hline\noalign{\smallskip}
$l_{1}l_{2}l_{3}$ & $m_{1}m_{2}$ & $|\phi\rangle_{C}$ & $R_{l_{1}l_{2}l_{3}m_{1}m_{2}}$\\
\noalign{\smallskip}\hline\noalign{\smallskip}
\multirow{4}{0.75cm}
{000} & 00 & $a|0\rangle+be^{i(\varphi_{1}+\varphi_{2})}|1\rangle$ & $I$\\
& 01 & $a|0\rangle-be^{i(\varphi_{1}+\varphi_{2})}|1\rangle$ & $Z$\\
& 10 & $a|0\rangle-be^{i(\varphi_{1}+\varphi_{2})}|1\rangle$ & $Z$\\
& 11 & $a|0\rangle+be^{i(\varphi_{1}+\varphi_{2})}|1\rangle$ & $I$\\
\hline
\multirow{4}{0.75cm}
{001} & 00 & $a|1\rangle+be^{i(\varphi_{1}+\varphi_{2})}|0\rangle$ & $X$\\
& 01 & $a|1\rangle-be^{i(\varphi_{1}+\varphi_{2})}|0\rangle$ & $ZX$\\
& 10 & $a|1\rangle-be^{i(\varphi_{1}+\varphi_{2})}|0\rangle$ & $ZX$\\
& 11 & $a|1\rangle+be^{i(\varphi_{1}+\varphi_{2})}|0\rangle$ & $X$\\
\hline
\multirow{4}{0.75cm}
{010} & 00 & $a|0\rangle+be^{i(\varphi_{1}+\varphi_{2})}|1\rangle$ & $I$\\
& 01 & $a|0\rangle-be^{i(\varphi_{1}+\varphi_{2})}|1\rangle$ & $Z$\\
& 10 & $a|0\rangle-be^{i(\varphi_{1}+\varphi_{2})}|1\rangle$ & $Z$\\
& 11 & $a|0\rangle+be^{i(\varphi_{1}+\varphi_{2})}|1\rangle$ & $I$\\
\hline
\multirow{4}{0.75cm}
{011} & 00 & $a|1\rangle+be^{i(\varphi_{1}+\varphi_{2})}|0\rangle$ & $X$\\
& 01 & $a|1\rangle-be^{i(\varphi_{1}+\varphi_{2})}|0\rangle$ & $ZX$\\
& 10 & $a|1\rangle-be^{i(\varphi_{1}+\varphi_{2})}|0\rangle$ & $ZX$\\
& 11 & $a|1\rangle+be^{i(\varphi_{1}+\varphi_{2})}|0\rangle$ & $X$\\
\hline
\multirow{4}{0.75cm}
{100} & 00 & $a|0\rangle+be^{i(\varphi_{1}+\varphi_{2})}|1\rangle$ & $I$\\
& 01 & $a|0\rangle-be^{i(\varphi_{1}+\varphi_{2})}|1\rangle$ & $Z$\\
& 10 & $a|0\rangle-be^{i(\varphi_{1}+\varphi_{2})}|1\rangle$ & $Z$\\
& 11 & $a|0\rangle+be^{i(\varphi_{1}+\varphi_{2})}|1\rangle$ & $I$\\
\hline
\multirow{4}{0.75cm}
{101} & 00 & $a|1\rangle+be^{i(\varphi_{1}+\varphi_{2})}|0\rangle$ & $X$\\
& 01 & $a|1\rangle-be^{i(\varphi_{1}+\varphi_{2})}|0\rangle$ & $ZX$\\
& 10 & $a|1\rangle-be^{i(\varphi_{1}+\varphi_{2})}|0\rangle$ & $ZX$\\
& 11 & $a|1\rangle+be^{i(\varphi_{1}+\varphi_{2})}|0\rangle$ & $X$\\
\hline
\multirow{4}{0.75cm}
{110} & 00 & $a|0\rangle+be^{i(\varphi_{1}+\varphi_{2})}|1\rangle$ & $I$\\
& 01 & $a|0\rangle-be^{i(\varphi_{1}+\varphi_{2})}|1\rangle$ & $Z$\\
& 10 & $a|0\rangle-be^{i(\varphi_{1}+\varphi_{2})}|1\rangle$ & $Z$\\
& 11 & $a|0\rangle+be^{i(\varphi_{1}+\varphi_{2})}|1\rangle$ & $I$\\
\hline
\multirow{4}{0.75cm}
{111} & 00 & $a|1\rangle+be^{i(\varphi_{1}+\varphi_{2})}|0\rangle$ & $X$\\
& 01 & $a|1\rangle-be^{i(\varphi_{1}+\varphi_{2})}|0\rangle$ & $ZX$\\
& 10 & $a|1\rangle-be^{i(\varphi_{1}+\varphi_{2})}|0\rangle$ & $ZX$\\
& 11 & $a|1\rangle+be^{i(\varphi_{1}+\varphi_{2})}|0\rangle$ & $X$\\
\noalign{\smallskip}\hline
\end{tabular}
\end{table}
Easy to see, whatever the outcome is, Charlie can achieve the desirable state by applying some reconstruction unitary operator $R_{l_{1}l_{2}l_{3}m_{1}m_{2}}$ (all of which are shown in Table 2), and the success probability in each cases is 1. Moreover, the probability of obtaining the state $|L_{l_{1}l_{2}l_{3}}\rangle_{B_{1}B_{2}C}$ for each of the outcomes $\{l_{1}l_{2}l_{3}\}$ is equal. So, the total success probability $P^{N\geq2}_{suc}$ can be calculated as below
\begin{equation}
{P^{N\geq2}_{suc}} = (\frac{1}{8} \times 1) \times 8   = 1.
\end{equation}

\section{Summary}
In this paper, through analyzing Bich et al.'s second protocol with $N>2$ preparers, we find that the success probability $P^{N>2}_{suc}=1/4 < 1$. In order to fix the drawback, we firstly constructed two sets of projective measurement bases: the real-coefficient basis $\{|u_{l_{1}l_{2}...l_{N}}\rangle{\kern 2pt}|l_{k}\in\{0,1\}\}$ and the complex-coefficient basis $\{|v_0\rangle_{B_j}, |v_1\rangle_{B_j}\}$, and further proposed an improved deterministic $N$-to-one JRSP protocol for an arbitrary single-qubit state with unit success probability (i.e, $P^{N\geq2}_{suc}=1$).

In addition, our protocol is also flexible and convenient: depending on the need, we can assign a receiver after sharing the quantum channel and the receiver may not be sufficiently well equipped. What is more, based on our protocol, we can structure a practical network: Alice can be regard as a Service Provider, she just distribute EPR pairs and measure her qubits; Other participants (Bob 1, Bob 2,\ldots, Bob N-1 and Charlie) can be regard as customers, while one of them is a receiver and the others are preparers. That is to say, with the help of Alice, arbitrary customer can be a receiver or preparer.

\begin{acknowledgements}
This work is supported by the National Nature Science Foundation of China (Grant Nos. 61103235, 61170321, 61373016 and 61373131), the Priority Academic Program Development of Jiangsu Higher Education Institutions (PAPD), the State Key Laboratory of Software Engineering, Wuhan University(SKLSE2012-09-41),  and the Practice Inovation Trainng Program Projects for the Jiangsu College Students (201310300018Z).
\end{acknowledgements}



\end{document}